\documentclass{aa}
\begin{document}
\outer\def\gtae {$\buildrel {\lower3pt\hbox{$>$}} \over 
{\lower2pt\hbox{$\sim$}} $}
\outer\def\ltae {$\buildrel {\lower3pt\hbox{$<$}} \over 
{\lower2pt\hbox{$\sim$}} $}
\newcommand{\ergscm} {ergs s$^{-1}$ cm$^{-2}$}
\newcommand{\ergss} {ergs s$^{-1}$}
\newcommand{\ergsd} {ergs s$^{-1}$ $d^{2}_{100}$}
\newcommand{\pcmsq} {cm$^{-2}$}
\newcommand{\ros} {\sl ROSAT}
\newcommand{\exo} {\sl EXOSAT}
\def\rchi{{${\chi}_{\nu}^{2}$}}
\newcommand{\Msun} {$M_{\odot}$}
\newcommand{\Mwd} {$M_{wd}$}
\def\Mdot{\hbox{$\dot M$}}

%
   \title{First {\sl XMM-Newton} observations of a Cataclysmic
   Variable I: Timing studies of OY Car\thanks{Based on observations
   obtained with XMM-Newton, an ESA science mission with instruments
   and contributions directly funded by ESA Member States and the USA
   (NASA).}}

   \author{Gavin Ramsay\inst{1}, Tracey Poole\inst{1}, Keith
   Mason\inst{1}, France C\'{o}rdova\inst{2}, 
William Priedhorsky\inst{3},
Alice Breeveld\inst{1}, Rudi Much\inst{4}, 
Julian Osborne\inst{5}, Dirk Pandel\inst{2}, Stephen Potter\inst{6},
Jennifer West\inst{2}, Peter Wheatley\inst{5}}

   \offprints{Gavin Ramsay: gtbr@mssl.ucl.ac.uk}

\institute{$^{1}$Mullard Space Science Lab, University College London,
Holmbury St. Mary, Dorking, Surrey, RH5 6NT, UK\\
$^{2}$Department of Physics, University of California, Santa
Barbara, California 93106, USA\\
$^{3}$Los Alamos National Laboratory, Los Alamos, NM 87545, USA\\
$^{4}$Astrophysics Division, ESTEC, 2200, AG Noordwijk, The
Netherlands\\
$^{5}$Department of Physics \& Astronomy, University of Leicester,
University Road, Leicester, LE1 7RH, UK\\
$^{6}$South African Astronomical Observatory, PO Box 9, Observatory
7935, South Africa}

\authorrunning{G. Ramsay et al} 
\titlerunning{{\sl XMM-Newton} timing studies of OY Car}

\date{}

\maketitle

\begin{abstract}

We present {\sl XMM-Newton} observations of the eclipsing, disc
accreting, cataclysmic variable OY Car which were obtained as part of
the performance verification phase of the mission. The star was
observed 4 days after an outburst and then again 5 weeks later when it
was in a quiescent state. There is a quasi-stable modulation of the
X-rays at $\sim$2240 sec, which is most prominent at the lowest
energies. We speculate that this may be related to the spin period of
the white dwarf. The duration of the eclipse ingress and egress in
X-rays is 20--30 sec. This indicates that the bulk of the X-ray
emission originates from the boundary layer which has a negligible
height above the surface of the white dwarf. The eclipse profile
implies a white dwarf of mass $M_{1}$=0.9--1.1\Msun\hspace{1mm} and a
secondary star of $M_{2}$=0.08--0.11\Msun\hspace{1mm}.


\end{abstract}

\section{Introduction}

Cataclysmic variables (CVs) are close binary systems in which the
secondary (usually a dwarf main sequence star) fills its Roche lobe
and transfers material onto the white dwarf primary. In non-magnetic
systems, this material forms an accretion disc around the primary.
Some of these show `dwarf nova' outbursts when the system brightens by
several magnitudes for a few days or weeks. These outbursts recur on a
timescale of weeks to months.

High inclination systems provide an opportunity to locate the X-ray
emitting region(s). Eclipses are not seen in the X-ray emission of OY
Car (Naylor et al 1988) nor in the EUV (Mauche \& Raymond 2000) during
outbursts, suggesting that the prime X-ray source (the boundary layer,
where the disc is decelerated and merges with the white dwarf) is
obscured at all phases. The residual X-ray flux is thought to be due
to scattering from an accretion disc wind. In contrast, during
quiescence X-ray eclipses have been seen in HT Cas (Wood et al 1995),
Z Cha (van Teeseling 1997) and OY Car (Pratt et al 1999) when the
boundary layer is not hidden by disc material. The duration of the
eclipse ingress/egress shows that the X-ray source is small and close
to the photosphere of the white dwarf, $<1.15R_{\rm wd}$ in HT Cas (Mukai
et al 1997).

X-ray observations of dwarf novae have been hampered by their
relatively low flux levels. Using instruments with large effective
areas, such as {\sl XMM-Newton} (Jansen et al 2001), we can examine
the eclipse profile of these systems with unprecedented precision and
accurately locate the source of the X-ray flux. Having an orbital
period of 1.51 hrs, OY Car was an ideal target for the performance
verification phase of {\sl XMM-Newton}. Specifically, OY Car was
chosen to test the timing capability of the {\sl XMM-Newton} Optical
Monitor (OM, Mason et al 2001).  This paper presents an initial
analysis of the optical and X-ray light curves. In a companion paper
(Ramsay et al 2001), we present the results of an analysis of the
X-ray spectra.

\section{Observations}

OY Car was observed using {\sl XMM-Newton} on 29--30 June 2000 and
again on 7 Aug 2000. Optical observations from the AAVSO (Mattei
2000a) indicate that OY Car was in outburst on 24--25 June (reaching
$\sim$12 mag) and had faded below $\sim$15 mag by June 27.
Observations taken using the OM indicate that $V\sim$16.0--16.2 mag on
June 29, indicating that it had faded rapidly after outburst. On Aug
7, OM observations gave $B\sim$15.7--15.9 mag. A series of 5 optical
spectra were also taken at the South African Astronomical Observatory
on Aug 7. The spectra imply $V\sim$15.8--16.5 and $B\sim$15.6--16.8,
indicating that, in the optical, OY Car was at a similar brightness
compared to the June observation.

Outbursts of OY Car vary in length. For those that last less than a
week, AAVSO observations (Mattei 2000b) show that the next outburst
typically occurs 180 days later, so it is unlikely that OY Car
experienced another outburst between the two {\sl XMM-Newton}
observations.

Data were collected in various observation modes of the OM. These
include $V$ band images and fast window mode data in $UVW1$
($\sim$2400--3400\AA) (all taken in June) and $B$ band images
($\sim$3800--4900\AA) and also Grism data (all taken in August).
Although OY Car was faint in both the UV and optical Grism
observations, an emission line is detected at H$\alpha$. Observations
were also made with each EPIC camera (PN, MOS1 \& MOS2, 0.1--12keV;
Turner et al 2001). There was a weak detection (0.05 ct s$^{-1}$
background subtracted) of OY Car in the RGS (0.3--2.1keV; den Herder
et al 2001).

The EPIC exposures were taken in full window mode using the medium
filter.  In both EPIC observations, the particle background increased
significantly towards the end. The data with high background were
therefore removed from the analysis. The resulting useful exposure and
the mean background subtracted count rates are shown in table 1: in
X-rays OY Car faded by a factor of 2.1 between the two observations.

In the June observations, 9 eclipses were observed in the EPIC cameras
and 2 in the OM UVW1 filter in fast mode. In the August observations,
2 and 3 eclipses were observed in the EPIC PN and MOS cameras
respectively and 3 eclipses in the OM B filter in fast window
mode. For the EPIC data we concentrate on the June observations since
the source was brighter and more eclipses were observed. The data were
processed using the {\sl XMM-Newton} Science Analysis System released
on 2000 July 12.

\begin{table}
\begin{tabular}{lrrr}
\hline
Date    & PN & MOS1 & MOS2\\
\hline
29/30 June 2000& 48ksec & 51ksec & 51ksec\\
orbit 102    & 1.09 ct/s& 0.43ct/s & 0.41ct/s\\
7 Aug 2000& 7ksec & 14ksec & 14 ksec\\
orbit 121  & 0.56 ct/s & 0.17 ct/s & 0.19 ct/s \\
\hline
\end{tabular}
\label{summary}
\caption{The duration of the EPIC observations and the mean background
subtracted count rates.}
\end{table}

\section{Light curves}

\subsection{Extracting the light curves}

Light curves were extracted from all 3 EPIC cameras using apertures
$\sim30^{''}$ in radius centered on OY Car, chosen so that the
aperture did not cover more than one CCD. This radius encompasses
$\sim$90 percent of the integrated PSF (Aschenbach et al
2000). Background lightcurves were extracted from the same CCD on
which the source was detected, scaled and subtracted from the source
light curves.

To increase the signal to noise ratio, the counts from the three EPIC
cameras were summed, having first corrected for absolute timing
discrepancies by cross-correlation and alignment of the sharp eclipse
ingress and egress. Similarly we extracted background subtracted light
curves from the OM fast mode data.  All our light curves were then
phased on the orbital period and aligned in such a manner that phase
$\phi\sim$0.0 was centered on the eclipse.

\subsection{General features}

\begin{figure*}
\begin{center}
\setlength{\unitlength}{1cm}
\begin{picture}(14,9.5)
\put(-4.,-1.6){\includegraphics{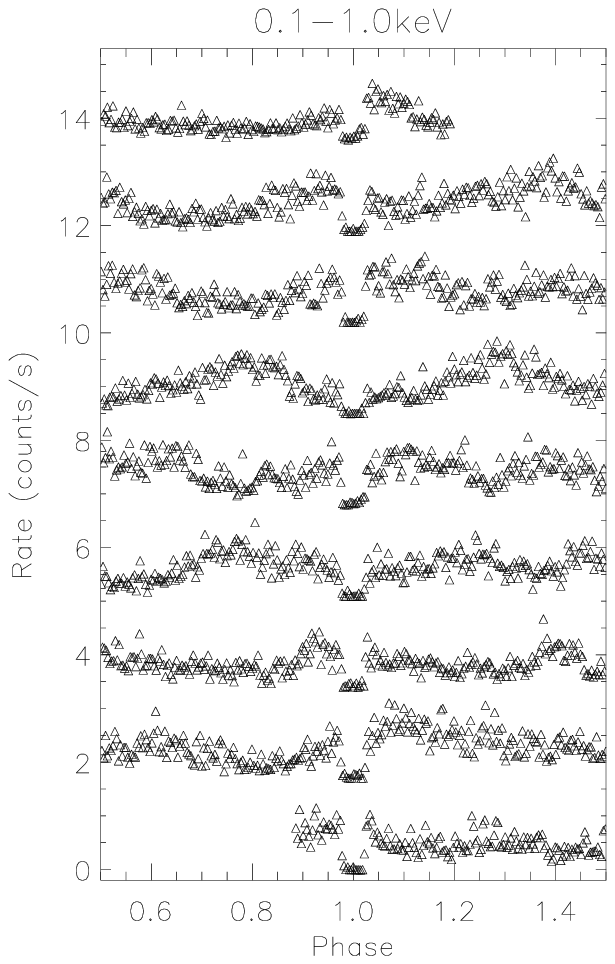}}
\put(1.8,-1.6){\includegraphics{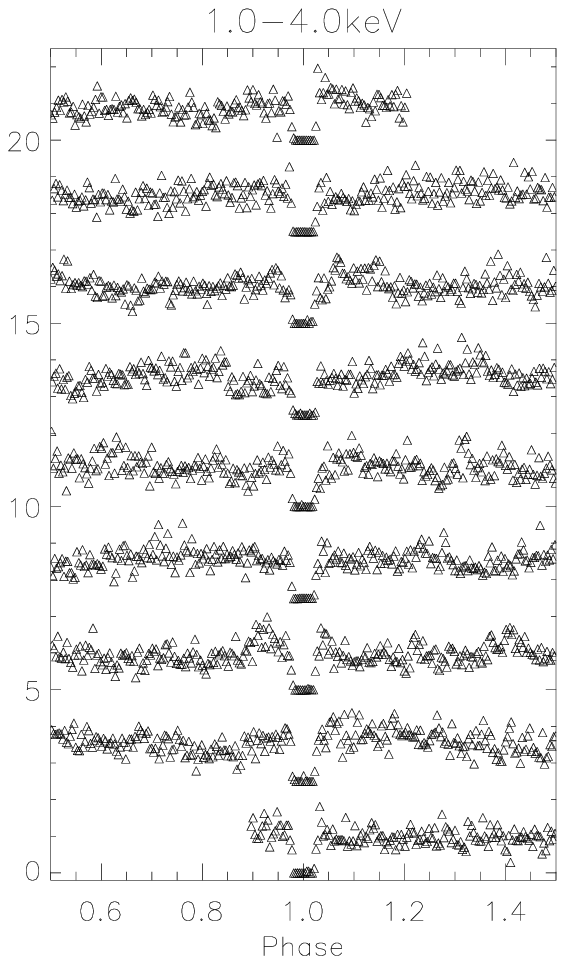}}
\put(8.,-1.6){\includegraphics{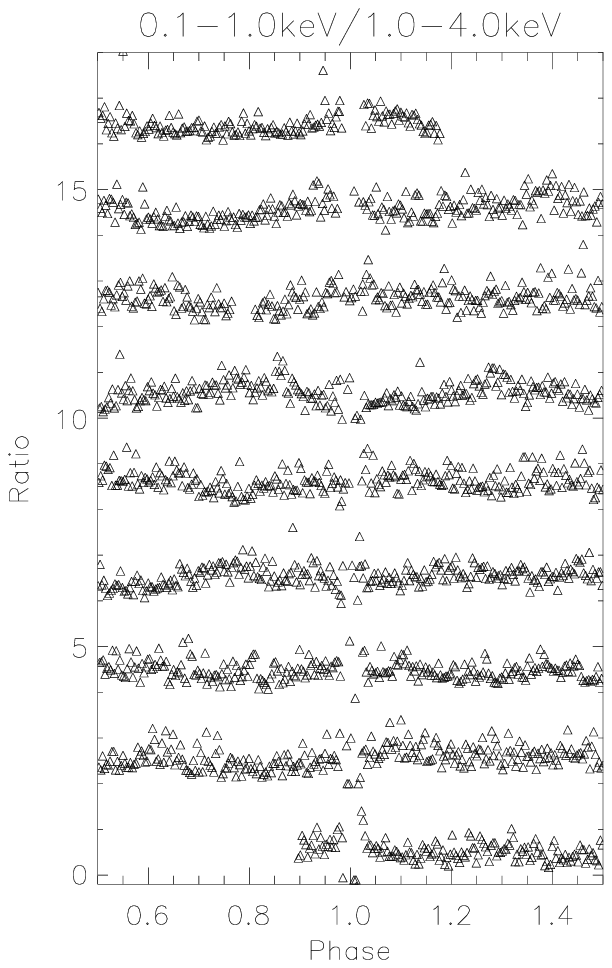}}
\end{picture}
\end{center}
\caption{The light curves obtained using the sum of data from all 3
EPIC cameras in the 0.1-1.0keV (left) and 1.0-4.0keV (center) energy
ranges and their ratio 0.1--1.0keV/1.0--4.0keV (right hand panel)
folded on the binary orbital period.  The light curves have been
binned into 10 sec bins and offset vertically by 1.7 ct/s, 2.5 ct/s
and 2.0 left to right respectively. Time increases from the bottom to
the top.}
\label{cycle} 
\end{figure*}

Figure \ref{cycle} shows the EPIC light curve, split into 0.1--1.0keV
and 1.0--4.0keV energy bands. We also plot the ratio of counts in the
0.1--1.0keV and 1.0--4.0keV bands. There is very little variation in
the 1.0--4.0keV band apart from the deep eclipses. In the soft band,
however, there are significant quasi-sinusodial variations, with
corresponding variations in spectra shape. These are not fixed in
orbital phase.

\subsection{Non-orbital modulations}
\label{2242}

\begin{figure}
\begin{center}
\setlength{\unitlength}{1cm}
\begin{picture}(14,10.5)
\put(-1.,-3.7){\includegraphics{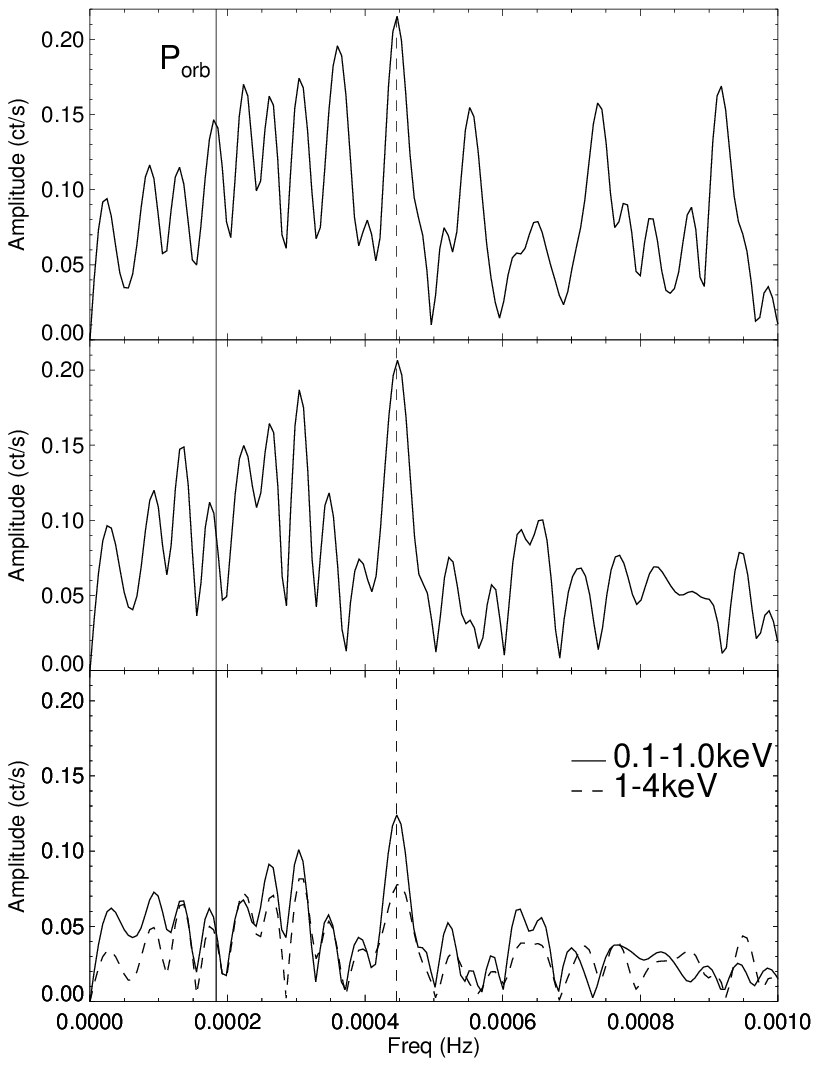}}
\end{picture}
\end{center}
\caption{Top panel: the amplitude spectrum of the background
subtracted summed EPIC light
curve (0.1--12.0keV), Middle panel: the amplitude spectrum 
with the removal of the average eclipse
profile (0.1--12keV), Bottom panel: as the middle
panel but for energy ranges 
0.1-1.0keV and 1.0-4.0keV.
The frequency of the orbital period corresponds
to P$_{\rm orb}$, while the dashed line corresponds to a period of 2240 sec.}
\label{power} 
\end{figure}

To explore the variations in the light curve further, we constructed a
Discrete Fourier Transform of the summed EPIC light curve.  We
show the amplitude spectrum in the energy range 0.1--12.0keV in the
top panel of Fig \ref{power}. This amplitude spectrum is
contaminated by harmonics of the orbital period. To subtract this
effect, we removed the eclipses from the light curve. The resulting
amplitude spectra (0.1--12.0keV, 0.1--1.0keV and 1.0--4.0keV) are
shown in the middle and lower panels of Figure \ref{power}.

The strongest amplitude peak occurs at 2240$\pm$60 sec (where the
error is a quarter of the full width of the amplitude peak at half
maximum), and this is most prominent at the lowest energies. The
orbital sidebands of the 2240 sec period are also present. However,
there are several other peaks which have similar amplitude
particularly at low frequencies. We also searched for much shorter
periods (less than 100 sec) and found no significant amplitude
peaks. As the OM fast mode data was short in duration it was not
possible to make a detailed search for periods on a timescale of a few
1000 sec in these data.

\begin{figure}
\begin{center}
\setlength{\unitlength}{1cm}
\begin{picture}(14,5.8)
\put(-2.,-14.7){\includegraphics{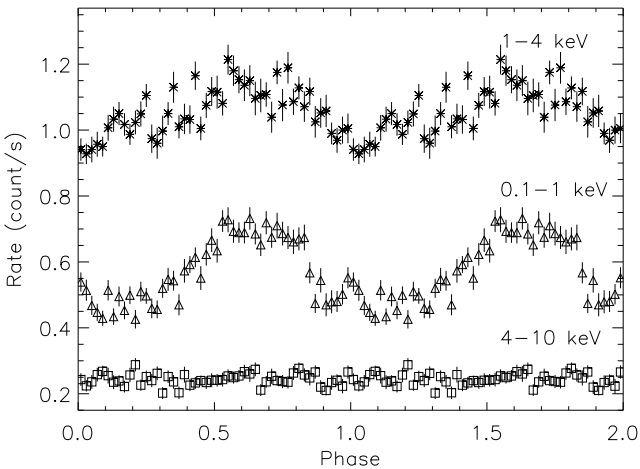}}
\end{picture}
\end{center}
\caption{The summed EPIC data with the eclipse profile removed and 
folded on the 2240 sec period found from our Discrete Fourier Transform.}
\label{operiod} 
\end{figure}

\begin{figure}
\begin{center}
\setlength{\unitlength}{1cm}
\begin{picture}(8,6)
\put(-2.2,-27.8){\includegraphics{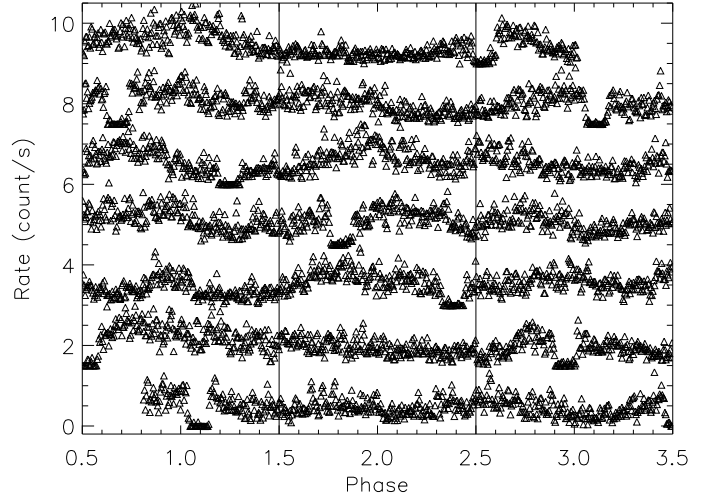}}
\end{picture}
\end{center}
\caption{The summed EPIC data (0.1--1.0keV) phased on the 2240 sec
period. Time increases from left to right and every 3 cycles the data
have been offset vertically by 1.5 ct/s.}
\label{ocycle} 
\end{figure}

The data folded on the 2240 sec period are shown in Figure
\ref{operiod}. This confirms the energy dependance of the modulation,
with an amplitude (maximum--minimum/the mean) of $\sim$43 per cent in
the 0.1--1.0keV band and $\sim$20 per cent in the 1.0--4.0keV band.
There is no evidence of a significant modulation in the 4.0--10.0keV
band. In Figure \ref{ocycle} we show the 0.1--1.0keV light curve
phased on the 2240 sec modulation. It is clear that if there is an
underlying modulation its amplitude is variable. However, all the
broad maxima appear centered on the phase range
$\phi_{\rm 2240}\sim$0.6--1.1. We discuss the possible origin of the 2240
sec period in \S \ref{discuss}.

\subsection{Multi-wavelength data}

\begin{figure}
\begin{center}
\setlength{\unitlength}{1cm}
\begin{picture}(12,11.)
\put(-1.3,-1.5){\includegraphics{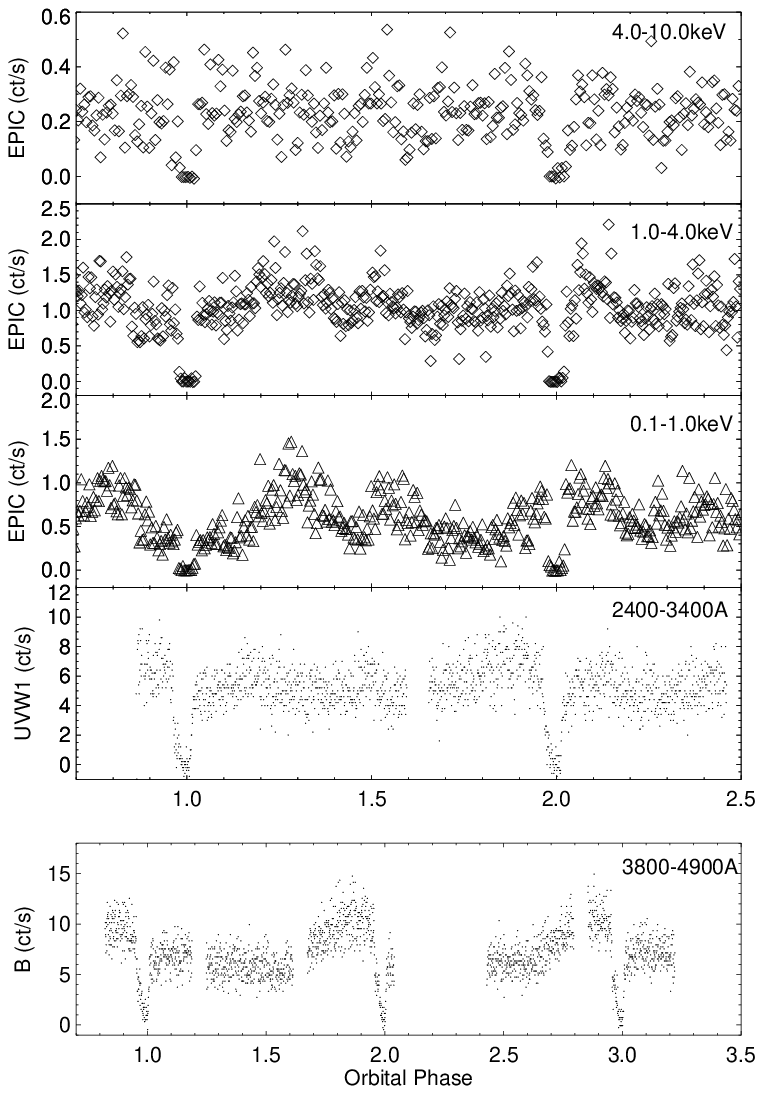}}
\end{picture}
\end{center}
\caption{Simultaneous EPIC (using the sum of the PN and MOS data) in
the 0.1-1.0 and 1.0--4.0keV energy bands and OM (UVW1,
$\sim$2400-3400\AA) light curves taken on 29 June 2000. The OM (B
filter) data were taken on 7 Aug 2000. The OM data
are binned into 5 sec bins and EPIC data into 20 sec bins apart from
the 4--10keV data which is 30 sec bins. Because of some uncertainty in
their absolute time, the light curves 
were phased on the orbital period then
aligned so that the eclipse was centered on $\phi$=0.0.}
\label{pnom} 
\end{figure}

We show in Figure \ref{pnom} simultaneous {\sl XMM-Newton} X-ray/UV
data which were obtained on June 29/30 2000.  These simultaneous data
cover 2 eclipses (eclipses 6 and 7 -- cf figure \ref{cycle}). We also
show in figure \ref{pnom} the $B$ filter (3800--4900\AA) fast mode
data taken in 7 August. Apart from the eclipses, there is no
variabilty in the UV data which corresponds to the low energy
variations seen in soft X-rays. For instance, the dip which is seen in
X-rays at $\phi\sim$1.4--1.5 is more prominent in the 0.1--1.0keV
energy band than at higher energies. However, the dip is not seen in
the UV data. One orbital cycle later, there is a dip in soft X-rays
between $\phi\sim$2.2--2.4. These dips maybe related to the absorption
features in the light curves of low mass X-ray binaries and in the
dwarf nova U Gem (Mason et al 1988). This phenomenon is thought to be
due to the accretion disc having a significant vertical extent at
certain azimuths.

\subsection{The eclipse profile}
\label{eprof}

In the optical the eclipse profile of disc accreting CVs is made up of
several components (eg Wood et al 1989). The eclipse of the white
dwarf is defined to be centered on $\phi$=0.0 and has a sharp ingress
and egress. The eclipse of the accretion disc is also symmetrical, but
the ingress and egress are much more gradual. On the other hand the
eclipse of the bright spot (where the accretion stream impacts with
the accretion disc) is centered later in the orbital cycle. In the
far-UV the white dwarf is the strongest component (Horne et al 1994).

\begin{figure}
\begin{center}
\setlength{\unitlength}{1cm}
\begin{picture}(8,13.5)
\put(-1.5,-1.5){\includegraphics{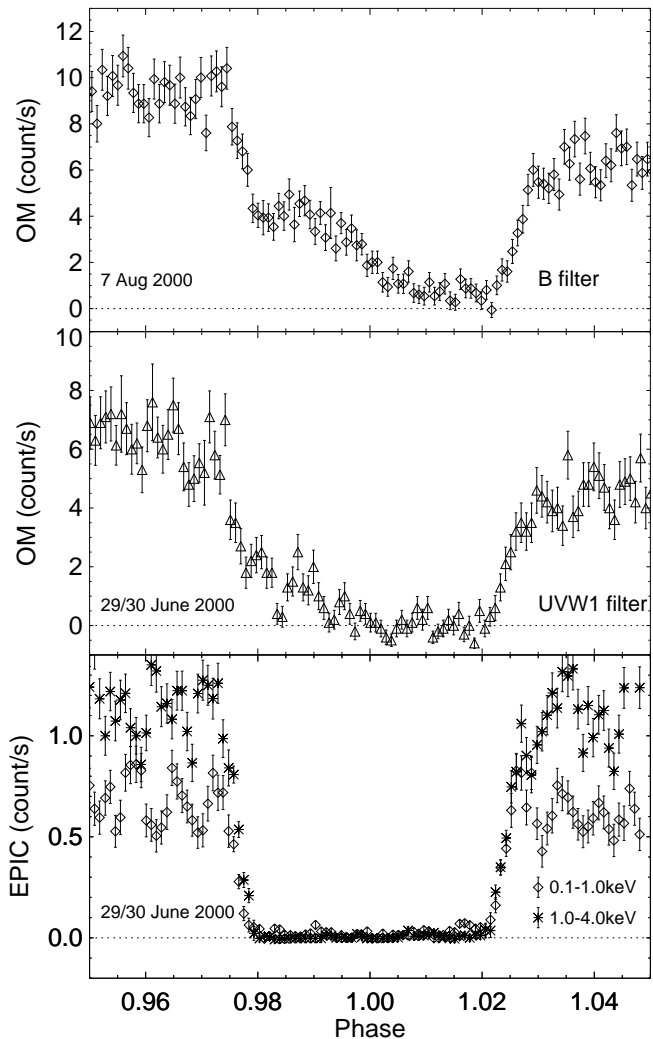}}
\end{picture}
\end{center}
\caption{Top panel: The eclipse profile in the OM $B$ filter
(3800--4900\AA). Middle panel: as above but the OM
UVW1 filter (2400--3400 \AA). Bottom panel: The eclipse profile
obtained using the sum of data from all 3 EPIC cameras in the
0.1-1.0keV and 1.0-4.0keV energy ranges. The data have been binned
into 5 sec bins.}
\label{eclipse} 
\end{figure}

In Figure \ref{eclipse} we show the mean eclipse profile in X-rays
(0.1--1.0keV and 1.0--4.0keV), the near UV (2400--3400\AA) and in B
(3800--4900\AA). The immediate point to note is the very rapid ingress
and egress in X-rays: 20--30 sec, which is roughly the expected time
to eclipse the diameter of the white dwarf. This suggests that
practically all the X-ray emission originates from the white
dwarf. Unlike magnetic CVs, which have relatively small accretion
regions and have ingress times $<$10sec (eg Steiman-Cameron \& Imamura
1999), the X-ray emission is not concentrated. Rather, the likely
source is the boundary layer which is extended in azimuth around the
white dwarf.  There is also some (rather weak) evidence that the
1.0--4.0keV flux is eclipsed sooner and comes out of eclipse later
than the 0.1--1.0keV flux.

Also of note is the short duration of the eclipse: defining the
eclipse duration as from first to fourth contact, we find the duration
is 290$\pm$10 sec in X-rays, 278sec$\pm$10sec in UVW1 and
286sec$\pm$10sec in B. We discuss the implications of these results in
\S \ref{discuss}. In the UVW1 filter the steep ingress/egress
into/from eclipse starts at the same orbital phase as in
X-rays. However, the flux does not return to its pre-eclipse flux --
this is due to the fact that the bright spot (which is visible in the
UVW1 band) is progressively foreshortened. In the B filter (which was
taken 5 weeks after the X-ray and UVW1 data), there is marginal
evidence that the steep ingress starts before the ingress in
X-rays. As in the UVW1 filter the flux does not return to its
pre-eclipse flux.

\subsection{The mid-eclipse flux}

An image made using EPIC data taken from times when the source was
eclipsed (defined as after 2nd contact and before 3rd) shows a faint
source at the location of OY Car. Using all three EPIC detectors, we
find a summed background subtracted count rate of 0.006$\pm$0.004
ct$^{-1}$ in the 1.0--12keV band and 0.027$\pm$0.004 ct$^{-1}$ in the
0.1--1.0keV band. This suggests that there was some residual low
energy X-ray flux during the eclipse.

To investigate this in greater detail we examined the source and
background light curves of each eclipse in the three separate EPIC
detectors. In eclipse 2, we found a small increase ($\sim$0.2 ct
s$^{-1}$) in the object flux for a very short time ($\sim$30 sec) in
the EPIC PN which was not seen in the MOS detectors. We also found the
object flux in eclipse 9 in the MOS 2 detector showed an increase
towards the end of the eclipse which was not seen in either of the
other detectors. We conclude that both these events were instrumental
in origin. Excluding these flux enhancements which we do not regard as
intrinsic to OY Car, we re-determined the background subtracted summed
count rate during the eclipse: 0.003$\pm$0.003 ct$^{-1}$
(1.0--12.0keV) and 0.024$\pm$0.004 ct$^{-1}$ (0.1--1.0keV).  The count
rate in the 1.0--12.0keV energy band is not significant, but the
0.1--1.0keV count rate is significant at the 6$\sigma$ level.

There are several possible origins of this emission. One is the
secondary star. Randich et al (1996) found that $L_{\rm X}/L_{\rm bol}$
saturates at $\sim1\times10^{-3}$ in late type stars, where $L_{\rm X}$ is
the luminosity in the 0.1--2.0keV energy band. For a dwarf secondary
star of $M_{2}$=0.1\Msun (\S \ref{mass}), and using the
mass-luminosity relationship of Malkov, Piskunov \& Shpil'kina (1997),
we find $log (L_{\rm bol}/L_{\odot})$=--2.96. This implies that
$L_{\rm X}<4\times10^{27}$ \ergss. To make a crude estimate of the
luminosity of OY Car during the eclipse, we determine the ratio of the
in-eclipse EPIC summed count rate over the mean EPIC summed count rate
(for 0.1--1.0keV $\sim$0.045). We then scale the luminosity in the
0.1--1.0keV energy band ($2\times10^{30}$ \ergss, Ramsay et al 2001)
using this ratio and derive $\sim9\times10^{28}$ \ergss for the
in-eclipse luminosity assuming a distance of 82pc (Wood et al
1989). Of course, the secondary star will have a softer spectrum than
the integrated spectrum of OY Car, so our derived luminosity may be
overestimated by a factor of $\sim$2.

Even taking this into account, we find that the observed luminosity
exceeds the maximum luminosity expected for a 0.1\Msun\hspace{1mm}
secondary by an order of magnitude. In order to match the estimated
luminosity during eclipse, a secondary of mass 0.3\Msun\hspace{1mm}
would be required (Malkov, Piskunov \& Shpil'kina 1997). Such a mass
is ruled out from our eclipse profile modelling (\S \ref{mass}). We
conclude that the weak detection of OY Car during eclipse is unlikely
to be due to the secondary star. An alternative scenario, is the weak
remnant of a large corona which is seen more prominently during
outburst (eg Naylor et al 1988). This seems to be more likely as an
outburst was observed 4 days before the {\sl XMM-Newton} observations.

\section{Discussion}
\label{discuss}

\subsection{The nature of the 2240 sec period}

The X-ray light curve of OY Car exhibits quasi-sinusoidal variability
that may have an underlying periodicity of $\sim$2240 sec. The
variation is most prominent at softer energies but has a variable
amplitude. Such a modulation could be due to energy dependent
absorption, perhaps in disk material circulating above the orbital
plane, or a quasi-stable bright spot on the photosphere of the white
dwarf.

The latter scenario would imply that the spin period of the white
dwarf ($P_{\rm spin}$) is $\sim$2240 sec. Marsh \& Horne (1998) observed
OY Car in the UV towards the end of a super-outburst and found
evidence for dwarf nova oscillations (17.94 and 18.15 sec). One
suggestion was that one of these periods represented the Keplerian
frequency of the inner edge of the accretion disc and the other a beat
frequency between the Keplerian frequency and $P_{\rm spin}$. This
scenario implies $P_{\rm spin}\sim$1500 sec.  Interestingly, a period of
1500 sec is close to the 1590-second orbital sideband of the 2240 sec
period that we find in the amplitude spectrum (Figure \ref{power}).

At this stage, we must regard this interpretation with a considerable
degree of caution. However, it is interesting to note that a spin
period of $P_{\rm spin}$=2240 sec, implies
$P_{\rm spin}/P_{\rm orb}$=0.41. The two Intermediate Polars (weakly magnetic
CVs) which have orbital periods below the period gap -- like OY Car --
(RX J1238--38 and EX Hya) have $P_{\rm spin}/P_{\rm orb}$=0.42 (Buckley et al
1998) and 0.68 (C\'{o}rdova, Mason \& Kahn 1985) respectively. If the
spin period in OY Car really is close to 2240 sec then it begs the
question, is OY Car an Intermediate Polar?

\subsection{The mass of the binary components}
\label{mass}

The most likely source of X-rays is from a boundary layer of
negligible height which is extended in azimuth around the white dwarf.
In \S \ref{eprof} we showed that in X-rays the ingress and egress of
the eclipse was very rapid taking 20--30 sec. In addition, the
duration of the eclipse in X-rays was 290$\pm$10 sec. Using the
observed times, we can place constraints on the system parameters.

There is good agreement regarding the system inclination: 83$^{\circ}$
(cf Wood et al 1989). There is, however, a wide variation in the
quoted mass of the white dwarf, $M_{1}$, varying from
0.33--1.26\Msun. The mass of the secondary star, $M_{2}$, is not much
better constrained, with estimates ranging from 0.07--0.15\Msun. There
is better agreement on the mass ratio, $q=(M_{2}/M_{1})$, where the
more recent estimates quoted in Wood et al (1989) gave $q=0.10-0.11$.

Using an inclination $i=83.3^{\circ}$ (Wood et al 1989) we determined
the ingress time and total eclipse duration for a range of
$M_{1},M_{2}$. We assume the Nauenberg (1972) mass-radius relationship
for a white dwarf and the secondary star is a main sequence star (ie
it is not over-massive for its spectral type as has been suggested for
some secondary stars).

We determine the size of the eclipsed source by tracing the Roche
potential out of the binary system along the line of sight from any
point in the vicinity of the white dwarf. We then determine the
ingress duration and eclipse duration for various combinations of
$M_{1},M_{2}$. Comparing these times with the observed times (\S
\ref{eprof}) we find that only for $M_{2}$=0.08--0.11\Msun\hspace{1mm}
and $M_{1}$=0.9--1.1\Msun\hspace{1mm} can we match the observed
times. Compared to previous estimates of the component masses these
are tightly constrained. If the boundary layer has a non-negligible
height, then $M_{1}$ will be more massive.

\end{document}